\documentclass{ws-ijmpb}
\usepackage{url}
\usepackage{graphicx}% Include figure files
\usepackage{dcolumn}% Align table columns on decimal point
\usepackage{bm}% bold math
\usepackage{wasysym}

\newcommand{\beq}{\begin{equation}}
\newcommand{\eeq}{\end{equation}}
\newcommand{\beqn}{\begin{eqnarray}}
\newcommand{\eeqn}{\end{eqnarray}}

\begin{document}

\markboth{Cenke Xu} {Unconventional Quantum Critical Points}

\catchline{}{}{}{}{}
%
%%%%%%%%%%%%%%%%%%%%%%%%%%%%%%%%%%%%%%%%%%%%%%%%%%%%%%%%%%%%%%%%%%%%

\title{Unconventional Quantum Critical Points
%\\
%USING \LaTeX\footnote{For the title, try not to use more than 3
%lines. Typeset the title in 10~pt Times Roman, uppercase and
%boldface.}
}

\author{Cenke Xu
%\footnote{Typeset names in
%8~pt Times Roman, uppercase. Use the footnote to indicate the
%present or permanent address of the author.}
}

\address{Department of Physics, University of California,
Santa Barbara, CA 93106, USA
%\footnote{State completely without abbreviations, the
%affiliation and mailing address, including country. Typeset in
%8~pt Times Italic.}\\
%firstauthor\_id@domain\_name\footnote{Typeset author e-mail
%address in single line.}
}

\maketitle

%\begin{history}
%\received{Day Month Year}
%\revised{Day Month Year}
%%\accepted{(Day Month Year)}
%%\comby{(xxxxxxxxxx)}
%\end{history}

\begin{abstract}

In this paper we review the theory of unconventional quantum
critical points that are beyond the Landau's paradigm. Three types
of unconventional quantum critical points will be discussed: (1).
The transition between topological order and semiclassical spin
ordered phase; (2). The transition between topological order and
valence bond solid phase; (3). The direct second order transition
between different competing orders. We focus on the field theory
and universality class of these unconventional quantum critical
points. Relation of these quantum critical points with recent
numerical simulations and experiments on quantum frustrated
magnets are also discussed.

\end{abstract}


\section{Introduction}

By definition, a critical point is associated with a continuous
phase transition between two different phases. In classical
systems, $i.e.$ systems at finite temperature where thermal
fluctuation dominates quantum fluctuation, a critical point is
almost always sandwiched between a high temperature thermal
disordered phase and a low temperature ordered phase where certain
global symmetry of the system is spontaneously broken, thus the
symmetry $H$ of the system at low temperature is a subgroup of the
symmetry $G$ at high temperature. This phase transition is
described by an order parameter $\Phi$ that carries a nontrivial
representation of $G$. The low temperature phase is characterized
by a nonzero expectation value $\langle \Phi \rangle$, which is
invariant under $H$. The low temperature phase is ``degenerate",
in the sense that the inequivalent states with the same free
energy form a manifold $\mathcal{M}$: \beqn \mathcal{M} = G/H.
\eeqn The critical point can be described by either a
Ginzburg-Landau (GL) theory formulated in terms of order parameter
$\Phi$, or by a ``Nonlinear sigma model" (NLSM) defined in
manifold $\mathcal{M}$.

As a simple example, let us consider a classical critical point of
a three dimensional magnet with a full spin SU(2) symmetry. In
this system, the full symmetry $G$ is SU(2). If we describe the
system using a GL theory, then the order parameter should be an
O(3) vector $\vec{\phi} = (\phi_1, \phi_2, \phi_3)$, and the GL
theory reads \beqn F = \int d^3 x \ \sum_{i = x}^z |\nabla_i
\vec{\phi}|^2 + r|\vec{\phi}|^2 + u (|\vec{\phi}|^2)^2.
\label{o3lsm} \eeqn In the disordered phase with $r > 0$, the
thermal expectation value $\langle \vec{\phi} \rangle = 0$, while
in the ordered phase with $r < 0$, the expectation value $\langle
\vec{\phi} \rangle \sim \vec{n} \neq 0$. Here $\vec{n}$ is a unit
vector: $|\vec{n}|^2 = 1$. In this case $H = U(1)$, which
corresponds to the spin rotation around $\vec{n}$. The manifold of
the ordered phase is $\mathcal{M} = G/H = SU(2)/U(1)$, which
corresponds to all the configuration of $\vec{n}$, and it is
equivalent to the two dimensional sphere $S^2$.

We can also describe this transition using a nonlinear sigma model
defined on manifold $S^2$. The manifold $S^2$ is parametrized by
the {\it unit} O(3) vector $\vec{n}$, with constraint $|\vec{n}|^2
= 1$. The NLSM reads \beqn F = \int d^3x \ \sum_{i = x}^z
\frac{1}{g} |\nabla_i \vec{n}|^2, \ \ \ |\vec{n}|^2 = 1.
\label{o3nlsm}\eeqn In Eq.~\ref{o3lsm}, the phase transition is
tuned by $r$, while in Eq.~\ref{o3nlsm} the phase transition is
tuned by $g$, $i.e.$ when $g < g_c$ ($g > g_c$) the system is in
an ordered (disordered) phase. Both theories Eq.~\ref{o3lsm} and
Eq.~\ref{o3nlsm} are supposed to describe a phase transition that
belongs to the 3D-O(3) Wilson-Fisher universality class. In order
to quantitatively show the equivalence between Eq.~\ref{o3lsm} and
Eq.~\ref{o3nlsm} at the critical point, one should compute the
critical exponents using renormalization group (RG) for both
models. Suppose we can compute the RG exactly, then it will tell
us the exact scaling dimension of $r$ in Eq.~\ref{o3lsm}, and the
exact scaling dimension of $\Delta g = g - g_c$ in
Eq.~\ref{o3nlsm} close to the critical point. These two scaling
dimensions should be identical.

A quantum critical point (QCP) is a continuous quantum phase
transition between two quantum ground states at zero temperature.
The formalism of classical critical point can be applied to many
QCPs, and these QCPs are called conventional QCPs. A conventional
QCP is usually sandwiched between an ordered phase with symmetry
breaking, and a disordered phase which is {\it gapped} and {\it
nondegenerate}. This disordered phase must be ``featureless",
namely by locally tuning the Hamiltonian, this phase can be
adiabatically connected to a fully gapped direct-product state
without any nontrivial correlation or entanglement, while the
system energy gap remains finite during this process.

%Since a direct product quantum state leads to trivial correlations
%between any operator, either local operator or nonlocal operator,
%the direct product quantum state is an analogue of the high
%temperature

The description of a conventional QCP is semiclassical, $i.e.$ it
is equivalent to a classical critical point. One can simply view
the time coordinate as one extra spatial coordinate, and write
down the GL theory or NLSM according to the symmetry. In this
formalism the trivial quantum disordered state is identified as
the thermal disordered high temperature phase. The only
complication here is the dimension of time and energy. In the
simplest case, the scaling dimension of time and energy is $-1$
and $1$ respectively, namely they have the same dimension as the
spatial coordinate and momentum, then in this case the effectively
GL theory and NLSM both have a Lorentz invariance. Thus a
$d-$dimensional conventional QCP is equivalent to a $D = d+1$
dimensional classical critical point, and all the computation
techniques that were applicable to classical critical points can
be straightforwardly generalized to the conventional QCPs.

This semiclassical formalism strongly relies on the nature of the
quantum disordered phase, $i.e.$ the semiclassical formalism is
only applicable when the disordered phase is completely trivial
(adiabatically connected to a direct product state). However, it
has been unambiguously shown that the ground states of many
quantum many-body systems have certain special nontrivial
structure called ``topological order"~\cite{wentopo}, even though
the spectrum of the system is gapped. With topological order, the
quantum disordered phase can no longer be adiabatically connected
to a direct product state, thus it is {\it inequivalent} to a
thermal disordered phase. Since the topological order cannot be
characterized using a semiclassical formalism, significant
modification should be made in our description.

The ordered phase of a quantum many-body system can also be
different from a classical system, although the most important
difference is usually encoded in its excitations instead of ground
state. To destroy an ordered phase, one usually has to proliferate
or condense the topological defects of the ordered phase. For
instance, to destroy a two dimensional superfluid phase, at finite
temperature the thermal fluctuation will proliferate the vortex
excitation, which leads to a Kosterlitz-Thouless
transition~\cite{kt}; at zero temperature, it is the vortex
condensation that destroys the superfluid phase. By definition a
topological defect will carry certain quantized topological
number. For example, the ground state manifold (GSM) $\mathcal{M}$
of a superfluid phase is $S^1$. Thus in superfluid a vortex defect
carries a quantized vorticity $2\pi \times$ Integer due to the
homotopy group $\pi_1[S^1] = Z$. However, in quantum systems,
sometimes a topological defect would carry some extra physical
quantum number, which is also quantized due to the quantization of
its topological number. Since the topological defect carries
physical quantum number, the condensation of the topological
defect will lead to another ordered phase with a different
symmetry breaking. For instance, it was shown by
Haldane~\cite{haldane1988} and Sachdev~\cite{sachdev1990} that the
Skyrmion defect of the two dimensional N\'{e}el order parameter of
a spin-1/2 system always carries lattice momentum, thus when the
Skyrmion of the N\'{e}el order condenses, the translation symmetry
of the lattice must be spontaneously broken. This type of quantum
phase transitions or QCPs are also unconventional.

In order to avoid confusions, in this paper we will consistently
distinguish two different concepts: topological defect
condensation $v.s.$ proliferation. Topological defect condensation
refers to the situation where topological defect are defined in
space only. This type of topological defects are usually referred
to as ``solitons", and they can be viewed as particles with their
own dynamics, and they can condense once their kinetic energy
becomes dominant. Proliferation refers to the situation where the
defects are defined in space-time, and these defects are usually
called ``instanton". Since these defects already live in
space-time, they can no longer be viewed as quantized particles,
but they will make nonzero contribution to the imaginary time path
integral, and this contribution can be either relevant or
irrelevant to the long wavelength continuum limit physics. When
this contribution becomes relevant, these topological defects
(instantons) ``proliferate".

Throughout this paper we will focus on continuous quantum phase
transitions only, because the usual wisdom is, when two states are
separated by a generic unfine-tuned continuous quantum phase
transition, then these two states indeed belong to two
``different" phases. However, if two states are separated by a
first order quantum phase transition, namely certain physical
quantity jumps discontinuously, then these two phases can still
belong to the same phase. For example, let us consider the
following GL theory for an Ising field $\Phi$: \beqn S = \int d^dx
d\tau \ (\partial_\mu\Phi)^2 + r \Phi^2 + g \Phi^3 + u \Phi^4 +
\cdots \eeqn Notice that this GL theory has a cubic term
$g\Phi^3$. With nonzero $g$, by simply minimizing this GL theory,
one can see that when $r$ is tuned to certain value, the
expectation value of $\Phi$ will jump discontinuously, thus there
is a first order transition. However, there is no qualitative
difference between these two states around this transition,
because they both have no symmetry left at all. Thus a first order
transition does not necessarily imply a qualitative change of the
state.

This paper is organized as follows: In section {\bf 2}, we will
discuss one example of conventional QCP, which is the QCP between
the Mott insulator phase and superfluid phase of the Bose-Hubbard
model. In section {\bf 3}, we will discuss the unconventional QCP
between an ordered phase and a topological phase, and we will take
the best-understood $Z_2$ topological phase as an example. In
section {\bf 4}, we will discuss the unconventional QCP between
two different ordered phases, for instance the QCP between
N\'{e}el and Valence Bond Solid (VBS) phase. In section {\bf 5}, a
unified field theory that contains all the unconventional QCPs
discussed in the previous sections will be discussed.

\section{An example of Conventional QCP}

The most well-known quantum critical point, is the QCP between the
Mott insulator (MI) and superfluid (SF) phase in the Bose-Hubbard
model. This model was first studied as a toy model in
Ref.~\cite{bosehubbard}, and later it was shown that this is
actually a perfect model to describe the spinless bosonic atoms
trapped in an optical lattice~\cite{greiner}.

The Bose Hubbard model reads \beqn H = \sum_{<i,j>} - t
b^\dagger_i b_j + H.c. + \frac{U}{2} (n_i  - \bar{n})^2.
\label{bhhamil}\eeqn This model has a global U(1) symmetry $b_i
\rightarrow \exp(i\theta) b_i$, which corresponds to the
conservation of the total boson number. The phase diagram of this
model is tuned by two parameters $\bar{n}$ and $t / U$. In the SF
phase, the expectation value $\langle b_i \rangle \neq 0$, and the
global U(1) symmetry of the model is spontaneously broken. When
$\bar{n}$ is an integer, the MI phase of this simple model is a
trivial quantum disordered phase, namely it is adiabatically
connected to a direct product state: $ \prod_i
(b^\dagger_i)^{\bar{n}}|0\rangle $. Thus in this model the SF-MI
transition is a conventional QCP, and it can be described
semiclassically.

\begin{figure}
\begin{center} \includegraphics[width=5.0 in]{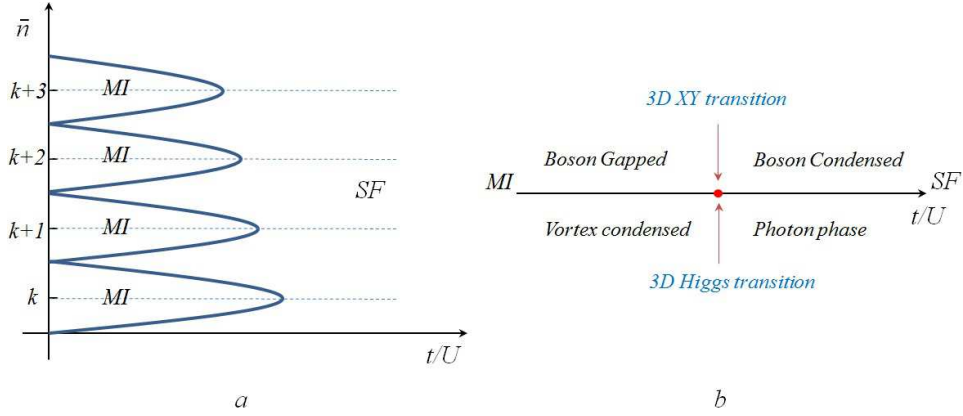}
\end{center}
\caption{{\it (a)}. The phase diagram of Eq.~\ref{bhfield}, tuned
by both $\bar{n}$ and $t/U$. There is an emergent Lorentz
invariance when $\bar{n}$ is an integer (dashed lines in this
phase diagram). {\it (b)}. Interpretation of the MI-SF transition
in terms of bosons, and in terms of vortices.} \label{QCP}
\end{figure}

In order to describe the MI-SF transition, we should first
introduce the continuum limit order parameter $\psi(x) \sim b_i$,
which carries the same representation of the global U(1) symmetry
as $b_i$. The MI-SF transition can be described by the following
field theory: \beqn S = \int d\tau d^dx \ \mu \psi^\ast
\partial_\tau \psi + |\partial_\tau \psi|^2 + \sum_i c^2 |\partial_i
\psi|^2 + r |\psi|^2 + g |\psi|^4 + \cdots \label{bhfield}\eeqn
The ellipses include other irrelevant terms allowed by symmetry.
When $\mu = 0$, the field theory Eq.~\ref{bhfield} has a
particle-hole symmetry $\psi\rightarrow \psi^\ast$. However, there
is no precise particle-hole (PH) symmetry for boson systems, thus
in the ellipses of Eq.~\ref{bhfield}, there are PH-symmetry
breaking terms like $\psi^\ast (\partial_\tau)^3 \psi + H.c.$.
This term breaks the PH symmetry of the field theory
Eq.~\ref{bhfield}, but it is irrelevant at the QCP $r = 0$. Thus
when $\mu = 0$, the PH symmetry becomes exact in the continuum
limit, where all the irrelevant terms flow to zero.

The mean field phase diagram of Eq.~\ref{bhfield} is apparently PH
symmetric at the lines $\bar{n} = k$~\cite{bosehubbard}, where $k
\in \mathrm{Integers}$. Thus close to the lines $\bar{n} = k$, we
can identify $\mu\sim \bar{n} - k$, $i.e.$ the system has an
emergent Lorentz invariance when $\bar{n} = k$. Let us focus on
the spatial dimension $d = 2$, then when $d = 2$ and $\mu = 0$
this QCP is precisely described by a classical three dimensional
GL theory with U(1) symmetry, thus this QCP belongs to the 3D O(2)
(or 3D XY) Wilson-Fisher universality class.

We can also describe this MI-SF transition in a different way. In
this phase diagram, SF is the ordered phase with global symmetry
breaking. Thus this transition can also be viewed as the
condensation of the topological defects of the SF phase, $i.e.$
the vortices of the SF phase. In order to describe this transition
in terms of vortices, we need to go to the dual picture. Inside
the superfluid phase, the low energy physics can be described by
the following rotor model: \beqn H = \int d^2x \
\frac{\tilde{u}}{2} (\delta \hat{n})^2 + \rho_s (\vec{\partial}
\theta)^2.  \eeqn where $\delta \hat{n} = \hat{n} - \bar{n}$ is
the density fluctuation above the average filling of the bosons;
$\theta$ is defined as $\psi \sim \sqrt{\rho_s} \exp(i\theta)$,
and $\rho_s$ is the superfluid stiffness. $\delta \hat{n}$ and
$\theta$ are a pair of conjugate variables, namely $[\delta
\hat{n}_x, \theta_{x^\prime}] = i \delta_{x,x^\prime}$.

The duality transformation is formulated as follows: define dual
vector field $\vec{E}$ and $\vec{A}$ as \beqn
\partial_i \theta = 2\pi \epsilon_{ij} E_j, \ \ \ \delta \hat{n} =
\frac{1}{2\pi} (\vec{\partial} \times \vec{A}), \eeqn the
commutation relation between $\delta \hat{n}$ and $\theta$
guarantees that $\vec{A}$ and $\vec{E}$ satisfy the algebra of a
pair of vector canonical variables: $[E^a_{x}, A^b_{x^\prime}] = i
\delta_{ab} \delta_{x, x^\prime}$. Also, since only the curl of
$\vec{A}$ is related to a physical quantity, the dual description
in terms of $\vec{A}$ and $\vec{E}$ must be invariant under the
following gauge transformation: $\vec{A} \rightarrow \vec{A} +
\vec{\partial} f$, which is the familiar gauge transformation for
U(1) gauge field. With the new variables, the rotor model
Hamiltonian is mapped to the Hamiltonian of a U(1) gauge
field:\beqn H = \int d^2x \ \frac{\tilde{u}}{8\pi^2}
(\vec{\partial} \times \vec{A})^2 + 4 \pi^2 \rho_s (\vec{E})^2.
\eeqn The Goldstone mode of the SF phase is dual to the photon of
the gauge field. In 2+1 dimension, an ordinary vector field has
two polarizations at each momentum $\vec{k}$. However, for a
vector gauge boson, one of the two polarizations is an unphysical
gauge degree of freedom, thus a gauge boson at 2+1d only has one
transverse physical mode at each momentum $\vec{k}$, this is why
it can be dual to a real scalar Goldstone mode.

%$i.e.$ {\it 2+1d gapless photon excitation is the Goldstone mode
%of the gauge flux condensate.}

This duality implies the following identity: \beqn \frac{1}{2\pi}
\epsilon_{ij} \partial_i \partial_j \theta = \partial_i E_i. \eeqn
The left side of this equation vanishes when $\theta$ is smooth in
the entire space, while it does not vanish when $\theta$ has a
singular vortex defect, $i.e.$ the vortex of $\theta$ is precisely
the gauge charge of the dual gauge field: $\partial_i E_i = n_v$,
here $n_v$ is the density of vortices. Inside the SF phase, an
isolated vortex has logarithmic divergent energy; in the dual
picture, an isolated gauge charge also has logarithmic divergent
energy due to its coupling to the dual U(1) gauge field. Thus the
dual theory of the SF phase is the following bosonic QED: \beqn S
= \int d\tau d^2x \ |(\partial_\mu - i A_\mu )\Phi|^2 + r_v
|\Phi|^2 + u |\Phi|^4 + \frac{1}{\tilde{e}^2} (F_{\mu\nu})^2.
\label{sfdual}\eeqn The complex field $\Phi$ is the vortex field,
$i.e.$ $\Phi(\vec{r})$ annihilate a vortex at position $\vec{r}$.
The SF-MI phase transition is driven by the condensation of the
vortices. In this theory, the phase with $r_v
> 0$ is an ``uncondensed" phase of vortex, and in this phase there
is one gapless photon, $i.e.$ the dual of the SF Goldstone mode.
On the other hand, in the phase with $r_v < 0$, the vortex
condenses, and the system is completely gapped due to the Higgs
mechanism, which is equivalent to the MI phase. Thus we can claim
that the 3D O(2) Wilson-Fisher critical point is dual to the
critical point of the bosonic QED.

%In this section, the vortex of the SF phase does not carry any
%physical quantity other than the quantized vorticity. Thus the
%vortex condensate is a trivial MI without any extra structure.
%Later we will also discuss situations where the vortex carries
%nontrivial conserved quantities, in which case the vortex
%condensation is an unconventional QCP.

\begin{figure}
\begin{center} \includegraphics[width=4.0 in]{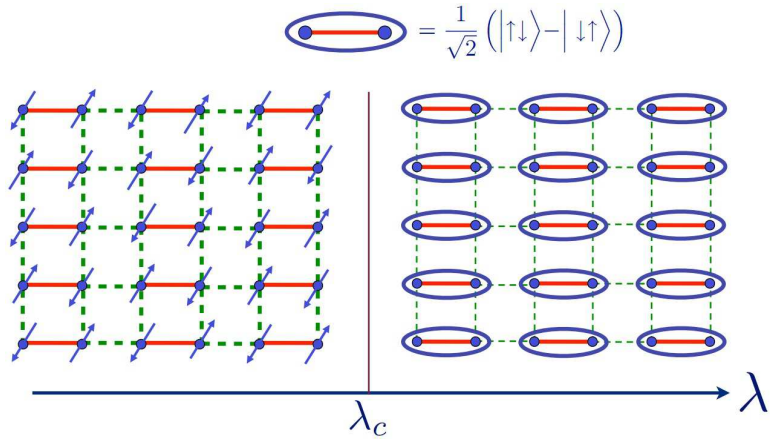}
\end{center}
\caption{ The phase diagram of the $J - \lambda$ model
Eq.~\ref{jjp}. The solid links and dashed links have Heisenberg
coupling $\lambda J$ and $J$ respectively. This QCP is a
conventional QCP that belongs to the 3D O(3) universality class. }
\label{Jlambda}
\end{figure}

Conventional QCPs also exist in some quantum spin models. However,
it was proved that for a spin-1/2 system with a local Hamiltonian,
the ground state has to be either gapless or gapped but
degenerate, thus a fully gapped nondegenerate direct product
ground state does not exist for an SU(2) invariant spin-1/2 model
on a lattice with one site per unit cell~\cite{LSM,hastings}. Thus
a conventional QCP {\it cannot exist} in ordinary spin-1/2 systems
on square or triangular lattices, unless the spin Hamiltonian
explicitly breaks the translation symmetry $i.e.$ the unit cell is
enlarged. For instance, let us investigate the following $J -
\lambda$ model (Fig.~\ref{Jlambda})~\cite{sachdevpic}: \beqn H =
\sum_{<\mathbf{i},\mathbf{j}> } \lambda J \vec{S}_\mathbf{i} \cdot
\vec{S}_\mathbf{j} + \sum_{<i, j> } J \vec{S}_i \cdot \vec{S}_j.
\label{jjp}\eeqn In this equation $<\mathbf{i}, \mathbf{j}>$
denotes solid links of the square lattice with Heisenberg coupling
$\lambda J$; $<i, j>$ denotes dashed links with Heisenberg
coupling $J $. When $\lambda \sim 1$, the system has an ordinary
N\'{e}el order; when $\lambda \gg 1$, the N\'{e}el order
disappears, and the system is disordered with a nondegenerate
ground state that is adiabatically connected to the following
direct product state: \beqn \prod_{<\mathbf{i}, \mathbf{j}>} |
\uparrow_{\mathbf{i}}\downarrow_{\mathbf{j}} -
\downarrow_{\mathbf{i}} \uparrow_{\mathbf{j}} \rangle. \eeqn This
is the exact ground state wave function when $\lambda = \infty$.
Notice that this state is a direct product between different unit
cells, while it is a maximally entangled state within one unit
cell.

The phase transition in Eq.~\ref{jjp} can be described by the
semiclassical GL theory Eq.~\ref{o3lsm} and NLSM Eq.~\ref{o3nlsm},
and it belongs to the O(3) Wilson-Fisher fixed point. If we start
with the N\'{e}el phase of this model, then this transition can be
viewed as condensing the Skyrmion defect of the N\'{e}el order
parameter. In this particular model, due to the explicit breaking
of the translation and rotation symmetry of the square lattice,
the Skyrmion will not carry any nontrivial physical quantum
number, thus the condensate of the Skyrmion is a featureless
direct product state. It is in fact a little surprising that one
has to cook up a relatively complicated spin model Eq.~\ref{jjp}
in order to realize a simple O(3) Wilson-Fisher transition in
quantum spin systems.

\section{Unconventional QCP between topological phase and ordered phase}

\subsection{$Z_2$ topological phase }

Starting with this section we will discuss unconventional QCPs,
and we will first discuss the QCP between semiclassical ordered
phases and phases with topological order. As we discussed in the
introduction section, when a phase has topological order, it can
no longer be adiabatically connected to a trivially gapped direct
product state, thus the semiclassical description needs
significant modifications.

In this section we will take the best understood $Z_2$ topological
phase in two spatial dimension as an example. The $Z_2$
topological phase is described by the $Z_2$ gauge field, which can
be obtained by spontaneously breaking a compact U(1) gauge
symmetry. The Hamiltonian of the compact U(1) gauge theory reads:
\beqn H = \sum_{j,\mu} - t\cos( \vec{\nabla} \times \vec{A} ) +
\frac{U}{2}E_{j,\mu}^2. \label{compactqed}\eeqn Both $A_{j, \mu}$
and $E_{j,\mu}$ are defined on links of a square lattice. A link
around site $j$ is denoted as $(j, \mu)$, with $\mu = \hat{x},
\hat{y}$. This Hamiltonian is always accompanied with the Gauss
law constraint: \beqn \sum_\mu \nabla_\mu E_\mu = \rho_j. \eeqn
$\rho_j$ is the local charge density on site $j$. The compact U(1)
gauge field $A_{j,\mu}$ is defined periodically: $A_{j,\mu} =
A_{j,\mu} + 2\pi$, and its canonical conjugate variable
$E_{j,\mu}$ must be discrete integers only.

In order to break the U(1) gauge symmetry to $Z_2$ gauge symmetry,
we can couple the compact U(1) gauge field to a U(1) rotor matter
field $\exp(i\phi)$: \beqn H = \sum_{\mu} - t \cos(\nabla_\mu \phi
- 2 A_\mu) + \cdots \label{phi} \eeqn The rotor field
$\exp(i\phi)$ can be viewed as a Cooper pair, which carries two
unit gauge charges. When $\phi$ is ordered, Eq.~\ref{phi} is
reduced to $ - t\cos(2A_\mu)$, which prefers $A_\mu$ to take only
two values $0 $ and $\pi$. When $t$ is strong enough, we can
effectively describe the physics of gauge field $A_\mu$ using the
following Ising variables $\sigma^z$ and $\sigma^x$ defined on the
links of the lattice: \beqn \sigma^z_{j,\mu} = \exp(iA_{j,\mu}), \
\ \sigma^x_{j,\mu} = \exp(i \pi E_{j,\mu}) = \cos(\pi E_{j,\mu}).
\label{sigma}\eeqn Here $A_{j,\mu}$ only takes two values $0$ and
$\pi$. Please note that $\sigma^z$ and $\sigma^x$ so defined
satisfy the ordinary Pauli matrix algebra.

Introducing $\sigma^z$ and $\sigma^x$ as Eq.~\ref{sigma}, the
Hamiltonian of compact QED in Eq.~\ref{compactqed} is reduced to
the following form: \beqn H_{z2} = \sum_{i} - K
\sigma^z_{i,x}\sigma^z_{i,y}\sigma^z_{i+y,x} \sigma^z_{i+x,y} -
h\sigma^x_{i,\mu}. \label{z2gauge1}\eeqn This Hamiltonian
Eq.~\ref{z2gauge1} has a special discrete symmetry: \beqn
\sigma^z_{i,\mu} \rightarrow \eta_i \sigma^z_{i,\mu} \eta_{i+\mu}.
\label{z2gaugetrans} \eeqn $\eta_i = \pm 1$ is an arbitrary $Z_2$
function on the lattice. Eq.~\ref{z2gaugetrans} is precisely the
$Z_2$ discrete gauge transformation. The model Eq.~\ref{z2gauge1}
is actually the minimal model that describes the $Z_2$ topological
phase, and the quantum $Z_2$ gauge theory.

Just like the U(1) gauge theory, the $Z_2$ gauge theory is always
subject to the following local gauge constraint: \beqn Q_{i} =
\sigma^x_{i,x}\sigma^x_{i-x,x}\sigma^x_{i,y}\sigma^x_{i-y,y} =
\chi_i. \eeqn Here $\chi_i = \exp(i\pi \rho_i)$. This $Z_2$ gauge
theory is called even or odd $Z_2$ gauge theory, when $\rho_i$ is
an even or odd integer. For example, if a $Z_2$ topological phase
is realized in a spin system, $\rho_i$ is usually the average
density of ``spinons" on every site: $\rho_i = \sum_\alpha
f^\dagger_{\alpha, i} f_{\alpha, i}$, and for spin-1/2 systems
there is precisely one spinon on every site, $i.e.$ $\rho_j = 1$.
Recently, it was demonstrated numerically that the $Z_2$
topological phase does exist in the $J_1 - J_2$ spin-1/2
Heisenberg model on the square lattice~\cite{hongchen,zhengcheng}.
Based on our analysis, this $Z_2$ topological phase must be an odd
theory.

In condensed matter systems, the $Z_2$ gauge symmetry is usually
obtained from spontaneously breaking a U(1) or even SU(2) gauge
symmetry by condensing a matter field which is the analogue of
``Cooper pair", and this gauge symmetry breaking occurs at a
rather high energy scale. At low energy we can safely ignore this
``Cooper pair", and describe everything using the effective $Z_2$
gauge theory.

When $K \gg h$ in Eq.~\ref{z2gauge1}, the system is in the
``deconfined phase" of the $Z_2$ gauge theory. In the ground state
of the deconfined phase, in addition to the constraint $Q_i =
\chi_i$, the ground state (approximately) satisfies
$\sigma^z_{i,x}\sigma^z_{i,y}\sigma^z_{i+y,x} \sigma^z_{i+x,y} =
1$ on every plaquette. In this deconfined phase, there are two
types of local excitations above the ground state: the first type
of excitation is a ``electric" excitation, or the $Z_2$ charge
excitation, which is a violation of the gauge constraint, $i.e.$
$Q_i = - \chi_i$ at some site $i$; the second type of excitation
is a ``magnetic" excitation, which corresponds to
$\sigma^z_{i,x}\sigma^z_{i,y}\sigma^z_{i+y,x} \sigma^z_{i+x,y} = -
1$ on certain plaquette. The magnetic excitations are usually
called the ``visons", and in terms of the original U(1) gauge
theory it is simply a local $\pi-$flux through one plaquette. The
unit electric and magnetic excitations satisfy the mutual semion
statistics, namely when a magnetic excitation adiabatically
encircles an electric excitation through a closed loop, the system
wave function will acquire a minus
sign~\cite{kitaev1997,kitaev2005}.

The ground state of the deconfined phase of the $Z_2$ gauge theory
is four fold degenerate on a torus. Starting with one of the
ground states, the other three ground states can be obtained by
inserting a vison ($\pi-$flux) through either hole of the torus.
This degeneracy is topological, in the sense that in the thermal
dynamical limit this degeneracy cannot be lifted through any weak
local perturbation in the Hamiltonian, even if this perturbation
breaks the $Z_2$ gauge symmetry.

Although the $Z_2$ topological phase is fully gapped, its
topological nature can be described by the following mutual
Chern-Simons field theory: \beqn S_{mcs} = \int d^2x d\tau \
\frac{i}{\pi}\epsilon_{\mu\nu\rho} a_\mu \partial_\nu b_\nu. \eeqn
$a_\mu$ and $b_\mu$ are two different U(1) gauge fields. This
mutual Chern-Simons (CS) theory leads to precisely four fold
degenerate ground states on a torus. Also, $a_\mu$ and $b_\mu$ are
minimally coupled to the currents of electric and magnetic
excitations of the $Z_2$ topological phase, and the mutual CS
theory guarantees that the electric and magnetic excitations see
each other as a $\pi-$flux, $i.e.$ they automatically have the
mutual semion statistics. Thus the mutual CS theory describes all
the key properties of a $Z_2$ topological phase.

Although the minimal model Eq.~\ref{z2gauge1} for the $Z_2$
topological phase looks quite abstract, the $Z_2$ topological
order can be very reliably realized in various (quasi-)realistic
models, such as the quantum dimer model on the triangular
lattice~\cite{sondhiz2}, a XXZ spin-1/2 model on the Kagome
lattice~\cite{balentsz2}, and also a quantum spin
Hall-Superconductor-Ferromagnet Josephson array~\cite{xufu}. For a
model that is not exactly soluble, the best way to verify the
$Z_2$ topological order is by computing the topological
entanglement entropy, which was introduced in
Ref.~\cite{wenentropy,kitaeventropy}.

A series of exactly soluble models have been constructed in
Ref.~\cite{fradkinz2}, and the phase diagram of these models have
both the $Z_2$ topological order and ordered phases with symmetry
breaking. In our paper we will focus on a more general discussion
about this type of transition, that is driven by the condensation
of topological excitations of this $Z_2$ topological state.

\subsection{QCP between $Z_2$ topological phase and superfluid}

Now let us assume that the $Z_2$ gauge field introduced in the
previous section is coupled to a bosonic matter field, and this
matter field carries a U(1) global quantum number in addition to
the $Z_2$ gauge charge. The simplest lattice model that describes
this physics reads \beqn H = \sum_{<j, \mu>} - t \sigma^z_{j, \mu}
\cos\left( \frac{\phi_j}{2} - \frac{\phi_{j + \mu}}{2}\right) +
\frac{U}{2} (n_j - \bar{n})^2 + H_{z2} \eeqn Here $\phi_j$ is the
phase angle of the boson matter field: $\Psi_j \sim
\exp(i\phi_j)$, and $\psi_j \sim \exp(i \phi_j /2)$ creates one
half of the boson. When $\phi$ is disordered, $i.e.$ $t$ is weak
compared with $U$, $\phi$ can be safely integrated out, and the
system is in the $Z_2$ topological phase; when $t$ is strong, or
$\phi$ is condensed, the system is in a SF phase.

Why do we couple the $Z_2$ gauge field to a half boson operator
instead of a single boson? The reason is that we want to make sure
that the condensate of $\phi$ is an ordinary superfluid whose
smallest vortex excitation has a $2\pi$ vorticity. In the
condensate of $\phi$, the smallest vortex is a $2\pi-$vortex of
$\phi$ ($\pi-$vortex of $\phi/2$), and it is bound with a vison
excitation of $Z_2$ gauge field $\sigma^z$. If the $Z_2$ gauge
field is coupled with a single boson $\exp(i\phi_j)$, then the
superfluid phase becomes a paired boson condensate, whose smallest
vortex has a $\pi-$vorticity.

As we discussed in the previous section, the SF-MI transition can
be interpreted as the condensation of the vortex of the SF phase.
Since the $2\pi-$vortex of the SF phase is bound with a vison, and
in the $Z_2$ deconfined phase the vison is a well-defined
excitation, thus the $Z_2$ topological phase is {\it not} a
condensate of the $2\pi-$vortex. Instead, the $Z_2$ topological
phase is a condensate of the $4\pi-$vortex, or the double-vortex
of the SF phase, which is not bound with any vison. This
transition driven by double-vortex condensation is usually
referred to as ``3D XY$^\ast$" transition.

Since the $Z_2$ gauge theory is fully gapped, it will not generate
any singular correlation for $\psi \sim \exp(i\phi/2)$ in the
infrared limit. Thus, this phase transition can be effectively
described by Eq.~\ref{bhfield}, although $\psi$ is not really a
gauge invariant operator. Let us focus on the case with $\mu = 0$,
where there is an emergent Lorentz invariance. In this case, this
transition belongs to the 3D XY universality class if we take
$\psi$ as an ``order parameter", namely the correlation length of
$\psi$ diverges as $\xi \sim r^{- \nu}$, and $\nu \sim 0.67$.
However, the scaling behavior of the physical order parameter
$\Psi \sim \psi^2$ is very different from the 3D XY universality
class. For example, let us consider the anomalous dimension
$\eta_{\Psi}$ of the physical order parameter $\Psi$, which is
defined as \beqn \Delta[\Psi] = (D - 2 + \eta_{\Psi})/2, \eeqn
where $\Delta[\Psi]$ is the scaling dimension of $\Psi$ at the
critical point. $\Psi$ corresponds to a bilinear composite field
at the 3D XY transition, thus $\Delta[\Psi] = \Delta[\psi^2]$. The
scaling dimensions of composite fields at a Wilson-Fisher critical
point have been calculated numerically with high
precision~\cite{vicari2003}, and quoting these results, we can
conclude that $\eta_{\Psi} = 1.49$ at this 3D XY$^\ast$
transition.

This anomalous dimension is enormous compared with the ordinary 3D
Wilson-Fisher transition. For instance, the ordinary 3D XY
transition has anomalous dimension $\eta \sim 0.03$, which is
orders of magnitude smaller than the 3D XY$^\ast$ transition. This
anomalous dimension can be verified numerically by computing the
scaling of the order parameter in the ordered phase close to the
critical point: \beqn \langle \Psi \rangle \sim r^{\beta}, \ \ \ \
\beta = \nu (D - 2 + \eta_{\Psi})/2. \eeqn The 3D XY$^\ast$
transition, along with its scaling dimensions have been confirmed
by quantum Monte Carlo simulation on a Hard-core boson model on
the Kagome lattice~\cite{melko}.

\subsection{QCP between $Z_2$ topological phase and spin order}

Recently a lot of efforts have been devoted to searching for spin
liquid phases in frustrated quantum spin models using various
numerical methods. So far it has been proposed that a fully gapped
spin liquid phase exists in the Kagome lattice spin-1/2
antiferromagnetic Heisenberg model~\cite{Jiang2008,white}, the
honeycomb lattice Hubbard model~\cite{meng}, and the $J_1-J_2$
spin-1/2 Heisenberg model on the square lattice~\cite{hongchen}.
In all these models, the numerical simulations have found a phase
without any symmetry breaking, and there is a finite gap for both
spinful and spin singlet excitations.

In this section we will only consider spin systems with a full
SU(2) symmetry (if the SU(2) spin symmetry is broken down to the
inplane U(1) symmetry, the situation reduces to the case discussed
in the previous section). Based on our understanding of spin
liquid state, when we see a fully gapped spin liquid state in
either experiments or numerical simulations, the first idea that
we have in mind is the $Z_2$ spin liquid, $i.e.$ the $Z_2$
topological phase. Then presumably in the $Z_2$ spin liquid phase
the electric excitation carries certain representation of the spin
SU(2) symmetry group, and the transition between the liquid phase
and the spin ordered phase is driven by the condensation of the
spin-carrying excitation (usually called spinon). Then the nature
of the spin order and the universality class of this transition
depend on the particular representation of this spin excitation.

The smallest representation of SU(2) is spin-1/2 representation,
and there is no consistent ``fractional" representation of SU(2)
group that is smaller than spin-1/2. Thus let us first assume the
spinon is a spin-1/2 boson, which is described by a two component
complex boson field $z_\alpha = (z_1, z_2)^t$, and $z_\alpha$ is
subject to the constraint $|z_1|^2 + |z_2|^2 = 1$. Just like the
previous section, $z_\alpha$ is coupled to a $Z_2$ gauge field in
the following way: \beqn H = \sum_{i, \mu}\sum_\alpha - t
\sigma^z_{i, \mu} z^\ast_{\alpha, i} z_{\alpha, i+\mu} + H.c. +
\cdots \label{o4field} \eeqn The condensed phase of $z_\alpha$ is
the spin ordered phase, while the disordered phase of $z_\alpha$
is the deconfined $Z_2$ topological phase.

Since $z_\alpha$ has in total two complex bosonic fields, $i.e.$
four real fields, then with the constraint $|z_1|^2 + |z_2|^2 =
1$, the entire configuration of $z_\alpha$ is equivalent to a
three dimensional sphere $S^3$. Since the spinon field $z_\alpha$
is coupled to a $Z_2$ gauge field, then the physical configuration
of the condensate of $z_\alpha$ is $S^3/Z_2$, which is
mathematically equivalent to group manifold SO(3). The
universality class of this transition is the 3D O(4)$^\ast$
transition, which is an analogue of the 3D XY$^\ast$ transition
discussed in the previous section. Since $z_\alpha$ itself is not
a physical observable, inside the condensate of $z_\alpha$ the
physical observables are the following three vectors: \beqn
\vec{N}_1 = \mathrm{Re}[z^t i\sigma^y \vec{\sigma} z], \ \
\vec{N}_2 = \mathrm{Im}[z^t i\sigma^y \vec{\sigma} z], \ \
\vec{N}_3 = z^\dagger \vec{\sigma} z. \eeqn A simple application
of the Fierz identity $\sum_a \sigma^a_{\alpha\beta}
\sigma^a_{\gamma\rho} = 2\delta_{\alpha\rho}\delta_{\beta\gamma} -
\delta_{\alpha\beta}\delta_{\gamma\rho}$ proves that these three
vectors are orthogonal with each other. At the 3D O(4)$^\ast$
quantum critical point, the anomalous dimension of $\vec{N}_i$ is
also very large: $\eta_{\vec{N}_i} \sim 1.37$~\cite{xuqi}.

One type of spin orders that has ground state manifold (GSM)
SO(3), is the noncollinear spin density wave (SDW), for instance
the standard 120 degree $\sqrt{3}\times \sqrt{3}$ SDW on the
triangular lattice, with order wave vector $\vec{Q} = (4\pi/3,
0)$. In this case, the vector $\vec{N}_i$ are defined as \beqn
\vec{S}(\vec{r}) &\sim& \vec{N}_1 \cos(\vec{Q}\cdot \vec{r}) +
\vec{N}_2 \sin(\vec{Q}\cdot \vec{r}), \cr\cr \vec{N}_3 &=&
\vec{N}_1 \times \vec{N}_2. \eeqn It is straightforward to check
that when vector fields $\vec{N}_1$ and $\vec{N}_2$ are ordered
uniformly on the lattice, $\vec{S}$ has the standard 120 degree
state on the triangular lattice. The 3D O(4)$^\ast$ QCP between
noncollinear SDW and $Z_2$ topological phase has been used to
explain the spin liquid phenomena observed in the organic
frustrated magnet $\kappa\mathrm{-(ET)_2 Cu_2
(CN)_3}$~\cite{xuqi}.

Since the first homotopy group of SO(3) is $\pi_1[\mathrm{SO(3)}]
= Z_2$, inside this spin ordered phase there are half-vortex
excitations, which are bound with the visons of the $Z_2$ gauge
field. Two of these half-vortices can annihilate each other.

Now let us assume the spin excitation of the $Z_2$ topological
phase carries a spin-1 representation. A spin-1 representation is
a vector representation of SU(2), $i.e.$ it can be parametrized as
a unit real vector $\vec{n}$, $ |\vec{n}|^2 = 1$. Now the coupling
between the spin excitation and $Z_2$ gauge theory reads \beqn H =
\sum_{i, \mu}\sum_a - t \sigma^z_{i, \mu} n^a_{i} n^a_{i+\mu} +
H.c. + \cdots \eeqn Again, since $\vec{n}$ couples to a $Z_2 $
gauge field, it is not a physical observable: $\vec{n}$ and
$-\vec{n}$ are physically equivalent. If vector $\vec{n}$
condenses, the condensate is in fact a spin nematic, or quadrupole
order, with quadrupolar order parameter \beqn Q^{ab} = n^a n^b -
\frac{1}{3} \delta_{ab} . \eeqn This spin order has manifold $S^2/
Z_2$, which also supports half-vortex excitations since
$\pi_1[S^2/Z_2] = Z_2$. The condensation transition of the vector
$\vec{n}$ belongs to the 3D O(3)$^\ast$ universality class.

In this section we have discussed two types of unconventional QCPs
between $Z_2$ liquid phase and spin orders. In either case, the
spin ordered phase is different from the ordinary collinear
N\'{e}el order, because a N\'{e}el order should have GSM $S^2$. In
particular, in both cases we have considered, the spin ordered
phase must have a nontrivial homotopy group $\pi_1$, which
corresponds to the vison excitation of the $Z_2$ topological
phase. In Ref.~\cite{hongchen} and Ref.~\cite{meng}, a {\it
continuous} quantum phase transition between a fully gapped spin
liquid phase and a N\'{e}el order was reported. If the fully
gapped spin liquid discovered in these numerical works is indeed a
$Z_2$ spin liquid as we expected, then such continuous quantum
phase transition is beyond our current understanding of
unconventional QCP.

\subsection{QCP between $Z_2$ topological order and VBS}

As we have mentioned, it has been proved that the ground state of
a spin-1/2 quantum magnet cannot be trivially gapped without any
degeneracy~\cite{LSM,hastings}. Thus if the ground state of a
spin-1/2 system has a short range spin-spin correlation, then
besides topological order, another possible scenario is the
valence bond solid (VBS) phase. The most naive picture of VBS
order is that, each spin forms a spin-singlet with one of its
neighboring spins, and these spin singlets form a crystal pattern
that breaks lattice symmetry, thus the ground state also has
degeneracy, although this degeneracy is due to spontaneous
symmetry breaking.

If there is a continuous quantum phase transition between the
$Z_2$ topological phase and the VBS phase, then this transition
can only be interpreted as the condensation of spinless
excitations of the $Z_2$ liquid phase, and this spinless
excitation must carry lattice momentum, in order to break the
lattice symmetry in its condensate. As we discussed in the
previous section, in the $Z_2$ liquid phase, the electric
excitations carry spin, then the only excitation that can drive
the transition into VBS is the magnetic excitation, or the vison.

In spin-1/2 $Z_2$ liquid phase, the $Z_2$ gauge theory is usually
{\it odd}. This is because the $Z_2$ gauge theory is subject to
the gauge constraint \beqn \prod_{\mathrm{links \ around \ site} \
i} \sigma^x_{ij} = (-1)^{\rho_i}, \label{honeyconstraint}\eeqn
where $\rho_i$ corresponds to the density of spinons on every
site, and in spin-1/2 systems, no matter we use bosonic or
fermionic spinons, $\rho_i$ is always 1.
%We can also consider constructing VBS ordered states as the%
%limit of ``hard dimers'', in which precisely one dimer (spin
%singlet) is attached to each site. The dimer constraint is
%translated into the Gauss law constraint in the gauge field
%language: $\vec{\nabla} \cdot \vec{E} = \eta_i $, and $\eta_i =
%\pm 1$ on two different sublattices. Now if the U(1) gauge
%symmetry is broken down to $Z_2$, we need to introduce a $Z_2$
%electric field $\sigma^x_{ij} = (-1)^{n_{ij}}$ on every link
%($n_{ij} = 0, 1$ denotes the absence and presence of dimer), and
%the gauge constraint becomes \beqn \prod_{\mathrm{links} \
%\mathrm{round} \ \mathrm{site} \ i} \sigma^x_{ij} = -1.
%\label{constraint}\eeqn
Let us consider the $Z_2$ gauge theory on
the honeycomb lattice first. With this $Z_2$ gauge constraint, we
can write down the simplest $Z_2$ gauge theory on the honeycomb
lattice as follows: \beqn H = \sum_{\hexagon} - K
\prod_{\mathrm{links} \ \mathrm{in} \ \hexagon}^6 \sigma^z_{ij} -
\sum_{i,j} h \sigma^x_{ij} + \cdots. \label{z2gauge}\eeqn The
first term is a sum of the ring product of the $Z_2$ gauge field
$\sigma^z_{ij}$ in every hexagon, and the second term is a $Z_2$
``string tension". The ellipses include other interaction terms
between $Z_2$ electric field.

\begin{figure}
\begin{center} \includegraphics[width=3.3 in]{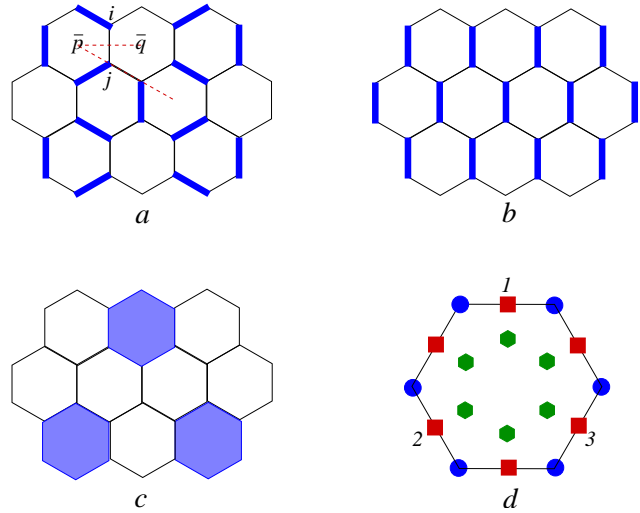}
\end{center}
\caption{({\it a}), $c-$VBS order. $\bar{p}$ and $\bar{q}$ are the
dual triangular lattice sites. We consider the nearest and 2nd
neighbor hopping for vison (vortex). ({\it b}), the $s-$VBS
pattern, realized when $h/8 < J < h$ in the dual Ising Hamiltonian
Eq.~\ref{jh}. ({\it c}), the four sublattice plaquette order,
realized when $w > 0$ in Eq.~\ref{cubic}. ({\it d}), the vison
(vortex) Brillouin zone. For weak 2nd neighbor vison (vortex)
hopping, the minima of band structure are located at the corner of
the BZ (circles); with intermediate 2nd neighbor hopping, there
are three inequivalent minima located at the center of the edges
of BZ (square); There are six inequivalent incommensurate minima
with strong 2nd neighbor hopping (hexagon). } \label{honeyvison1}
\end{figure}

When the $K$ term dominates everything else in Eq.~\ref{z2gauge},
the system is in the deconfined phase of the $Z_2$ gauge theory,
with topological degeneracy. When $h$ or other interaction terms
between $\sigma^x$ dominate $K$, the system enters the confined
phase. In order to analyze the confined phase, it is convenient to
go to the dual picture of the $Z_2$ gauge theory. Dual variables
$\tau^z$ and $\tau^x$ are defined on the dual lattice sites
$\bar{m}$, which are located at the center of the hexagons
(Fig.~\ref{honeyvison1}$a$): \beqn && \sigma^x_{ij} = -
\tau^z_{\bar{p}}\tau^z_{\bar{q}}, \ \ \bar{p} \ \mathrm{and} \
\bar{q} \ \mathrm{share} \ \mathrm{link} \ ij, \cr\cr &&
\prod_{\mathrm{links} \ \mathrm{around} \ \bar{p}}^6 \sigma^z_{ij}
= \tau^x_{\bar{p}}. \label{dual}\eeqn Introduction of
$\tau^z_{\bar{i}}$ automatically solves the odd $Z_2$ gauge
constraint Eq.~\ref{honeyconstraint}. Now the Hamiltonian becomes
an {\sl antiferromagnetic} transverse field Ising model on the
dual triangular lattice: \beqn H = \sum_{\bar{p}} - K
\tau^x_{\bar{p}} + \sum_{\bar{p},\bar{q}} J_{\bar{p},\bar{q}}
\tau^z_{\bar{p}}\tau^z_{\bar{q}} \eeqn For nearest neighbor sites
$\bar{p},\bar{q}$, $J_{\bar{p},\bar{q}} = h$. When
$J_{\bar{p},\bar{q}}$ dominates $K$, $\tau_{\bar{p}}^z$ takes on a
non-zero expectation value forming some pattern which optimizes
the $J_{\bar{p},\bar{q}}$ term. The non-zero ``condensate'' of
$\tau^z$ signals that the $Z_2$ gauge theory has entered the
confined phase.

The pattern of order in $\tau_{\bar{p}}^z$ depends upon the
detailed form of $J_{\bar{p},\bar{q}}$.  This can be analyzed by
treating $\tau^z_{\bar{p}}$ as a ``soft'' scalar field taking all
possible real values, rather than the integers $\pm 1$; this
approximation describes well the critical region in which
fluctuations on short time scales render the average of $\tau^z$
non-integral. Then, the quadratic form defined by
$J_{\bar{p},\bar{q}}$ can be diagonalized in wavevector space and
generically has multiple minima in its Brillouin zone.  Physically
the eigenvalues of this quadratic form define the dispersion
relation of visons in the $Z_2$ phase.  On entering the confined
phase, the location of these minima determines the VBS pattern.
Notice that the physical VBS order parameter should always be a
bilinear of $\langle \tau^z \rangle $, since under transformation
$\tau^z \rightarrow - \tau^z$ the physical quantity $\sigma^x$ is
unchanged. In the following we will discuss four types of VBS
patterns on the honeycomb lattice.

\subsubsection{$c-$VBS order on the honeycomb lattice}

Now let us take the simplest case, with nonzero
$J_{\bar{p},\bar{q}}$ only between nearest neighbor dual sites
$\bar{p},\bar{q}$. Taking $h > 0$, the model becomes the nearest
neighbor frustrated quantum Ising model with transverse field.
This model was studied in Ref.~\cite{sondhi2001a}. Solving the
band structure of $\tau^z$, we find two inequivalent minima at the
corners of the vison BZ: $\vec{Q} = (\pm \frac{4\pi}{3}, 0)$.
Expanding $\tau^z$ at these two minima, we obtain a complex local
order parameter $\psi$: \beqn \tau^z \sim \psi
e^{i\frac{4\pi}{3}x} + \psi^\ast e^{- i\frac{4\pi}{3}x}.
\label{complex1} \eeqn The low energy physics of visons should be
fully characterized by $\psi$.

Under discrete lattice symmetry, $\psi$ transforms as \beqn T_1
&:& x \rightarrow x + 1, \ \ \psi \rightarrow
e^{i\frac{4\pi}{3}}\psi, \cr\cr T_2 &:& x\rightarrow x +
\frac{1}{2}, \ y \rightarrow y + \frac{\sqrt{3}}{2}, \ \ \psi
\rightarrow e^{i\frac{2\pi}{3}}\psi, \cr\cr \mathrm{P}_y &:& x
\rightarrow -x, \ \ \psi \rightarrow \psi^\ast, \cr\cr
\mathrm{P}_x &:& y \rightarrow - y, \ \ \psi\rightarrow \psi,
\cr\cr \mathrm{T}  &:& t \rightarrow -t, \psi \rightarrow
\psi^\ast, \cr\cr \mathrm{R}_{\frac{2\pi}{3}} &:& \psi \rightarrow
\psi. \label{transformc}\eeqn $\mathrm{R}_{\frac{2\pi}{3}}$ is the
rotation by $2\pi/3$ around the center of hexagon.

The transformations in Eq.~\ref{transformc} determine that the low
energy Lagrangian for $\psi$ reads \beqn \mathcal{L} =
|\partial_\mu \psi|^2 + r |\psi|^2 + u|\psi|^4 + w (\psi^6 +
\psi^{\ast 6}), \label{XY} \eeqn $i.e.$ The condensation of $\psi$
is described by a $3D$ XY transition with $Z_6$ anisotropy, which
is an irrelevant perturbation at the $3D$ XY universality class.
The physical VBS order parameter $V$ should be a bilinear of
$\psi$, $i.e.$ $V \sim \psi^2$. It is straightforward to check
that $V$ transforms in the same way as the columnar VBS ($c-$VBS)
order parameter on the honeycomb lattice. Thus more precisely,
this transition belongs to the 3D XY$^\ast$ universality class,
where the anomalous dimension VBS order parameter $V$ is $\eta_{V}
\sim 1.49$. Notice that on the honeycomb lattice the $c-$VBS and
the $\sqrt{3}\times \sqrt{3}$ plaquette order have the same
symmetry, hence the condensate of $\psi$ can be either the $c-$VBS
or the plaquette order depending
on the sign of $w$. %Recently this plaquette order has been
%observed with exact diagonalization on frustrated spin models on
%the honeycomb lattice \cite{plaquette}. The $Z_6$ anisotropy
%introduced by the $w$ term in Eq.~\ref{XY} is an irrelevant
%perturbation at the 3d XY fixed point.

%{\bf should
%one comment on the other possible pattern for the opposite sign of
%$w$ in the honeycomb case?}

If we approach this transition from the $c-$VBS side of the phase
diagram, this transition can be interpreted as a proliferation of
the vortex of $\psi$ $i.e.$ double vortex of VBS order paramter
$V$, while the single vortex of $V$ is still gapped. In fact, the
single vortex core of the $c-$VBS is attached with a spinon
(analogous to the square lattice case discussed in
Ref.~\cite{levinsenthil}), condensation of single vortex will lead
to a spinon condensate, which corresponds to certain spin order.
However, if the spinon gap is finite, the finite temperature
thermal fluctuation can proliferate the single vortex. Therefore
although the quantum phase transition is driven by double
vortices, the finite temperature phase transition is still driven
by single vortex, hence at finite temperature the $Z_6$ anisotropy
of Eq.~\ref{XY} becomes the $Z_3$ anisotropy, and there is no
algebraic Kosterlitz-Thouless phase at finite temperature. This is
a key difference between our current case and a physical
transverse field frustrated quantum Ising model, where a finite
temperature algebraic phase is expected \cite{sondhi2001a}.
%{\bf
%is
%  there some peculiarity reflecting the crossover from double to
%  single vortices at low but non-zero $T$?  What is the measureable
%  consequence here?}

\subsubsection{$s-$VBS order and four-fold plaquette order on the
honeycomb lattice}

Now we modify the $Z_2$ gauge theory in Eq.~\ref{z2gauge} by
turning on the interaction between $Z_2$ electric field $\sigma^x$
on second nearest neighbor links: \beqn H_J = \sum_{\mathrm{2nd \
neighbor \ links}} J \sigma^x_{ij} \sigma^x_{kl}. \eeqn In the
dual theory this electric field interaction becomes a next nearest
neighbor hopping of $\tau^z$, and the full dual Hamiltonian reads
\beqn H = \sum_{\bar{p}} - K \tau^x_{\bar{p}} + \sum_{<
\bar{p},\bar{q} > } h \tau^z_{\bar{p}}\tau^z_{\bar{q}} + \sum_{\ll
\bar{p},\bar{q} \gg } J \tau^z_{\bar{p}}\tau^z_{\bar{q}}.
\label{jh} \eeqn The vison minima $(\pm \frac{4\pi}{3}, 0)$ are
stable with $J / h < 1/8$. When $1/8 < J / h < 1$, the minima of
the vison band structure are shifted to three inequivalent points
on the edges of BZ (Fig.~\ref{honeyvison1}$d$): \beqn \vec{Q}_1 =
(0, \frac{2\sqrt{3}\pi}{3}), \ \ \ \vec{Q}_2 = ( - \pi, -
\frac{\sqrt{3}\pi}{3}), \ \ \ \vec{Q}_3 = ( \pi, -
\frac{\sqrt{3}\pi}{3}). \label{honeysvbsq}\eeqn Notice that $-
\vec{Q}_a$ are equivalent to $\vec{Q}_a$ in the BZ.

Now three low energy modes can be defined by expanding $\tau^z$ at
momenta $\vec{Q}_a$: \beqn \tau^z \sim \sum_a \varphi_a \
e^{i\vec{Q}_a \cdot \vec{r}}. \label{complex2} \eeqn Since
$\vec{Q}_a$ and $- \vec{Q}_a$ are equivalent, all three fields
$\varphi_a$ are real. Under lattice symmetry, $\varphi_a$
transform as \beqn T_1 &:& \varphi_1 \rightarrow \varphi_1, \ \
\varphi_{2},\varphi_{3} \rightarrow - \varphi_{2}, -\varphi_{3},
\cr\cr T_2 &:& \varphi_{1},\varphi_{2} \rightarrow -
\varphi_{1},-\varphi_{2}, \ \ \varphi_3 \rightarrow \varphi_3,
\cr\cr \mathrm{P}_y &:& \varphi_1 \rightarrow \varphi_1, \ \
\varphi_2, \varphi_3 \rightarrow \varphi_3, \varphi_2, \cr\cr
\mathrm{P}_x &:& \varphi_1 \rightarrow \varphi_1, \ \ \varphi_2,
\varphi_3 \rightarrow \varphi_3, \varphi_2, \cr\cr \mathrm{T} &:&
\varphi_a \rightarrow \varphi_a, \cr\cr
\mathrm{R}_{\frac{2\pi}{3}} &:& \varphi_1 \rightarrow \varphi_2, \
\varphi_2 \rightarrow \varphi_3, \ \varphi_3 \rightarrow
\varphi_1. \label{transforms}\eeqn

Now the symmetry allowed Lagrangian for $\varphi_a$ up to the
quartic order reads \beqn \mathcal{L} = \sum_{a} (\partial_\mu
\varphi_a)^2 + r \varphi_a^2 + u (\sum_a \varphi_a^2)^2 + w
(\sum_a \varphi_a^4 ). \label{cubic}\eeqn This is an O(3) model
with cubic anisotropy. There are two possible types of condensates
of $\varphi_a$:

({\it i}) When $w > 0$, the condensate $\langle \vec{\varphi}
\rangle$ are along the diagonal directions, and there are in total
four independent states with $\langle \vec{\varphi} \rangle \sim
(1, 1, 1)$, $(-1, -1, 1)$, $(-1, 1, -1)$, $(1, -1, -1)$. According
to the transformation of $\vec{\varphi}$, these four states
correspond to the four-sublattice plaquette phase
(Fig.~\ref{honeyvison1}$c$).

({\it ii}) When $w < 0$, the condensate
$\langle\vec{\varphi}\rangle$ has three fold degeneracy: $\langle
\vec{\varphi} \rangle \sim (1, 0, 0)$, $(0, 1, 0)$ and $(0, 0,
1)$. These three condensates break the rotation symmetry of the
lattice, but they do not break the translation symmetry. This is
again because physical order parameters are bilinears of
$\varphi_a$, hence they are insensitive to the sign change of
$\varphi_a$ under translation. These three states correspond
precisely to the three staggered VBS ($s-$VBS) pattern
(Fig.~\ref{honeyvison1}$b$). Unlike the $c-$VBS, the $s-$VBS is no
longer described by an XY order parameter, and the phase
transition is not driven by vortex-like VBS defect.

The universality class of the QCPs described by Eq.~\ref{cubic}
was discussed carefully in Ref.~\cite{xubalents}.

%The universality class of Eq.~\ref{cubic} was studied extensively
%with $\epsilon = 4 - d$ expansion \cite{vicari2003}. The results
%is that the O(3) Heisenberg fixed point is not stable. For $w > 0$
%the transition is controlled by a stable cubic fixed point with
%nonzero fixed point values $w^\ast$ and $u^\ast$.  This case
%corresponds to the transition between the $Z_2$ spin liquid and
%the four-sublattice plaquette phase described above.  For $w < 0$,
%which corresponds to the transition to the $s-$VBS phase, there is
%instead a run-away flow, which most likely implies a first order
%transition. But if $w$ is small enough, for numerical simulations
%on finite system the transition between $Z_2$ spin liquid and
%$s-$VBS will be similar to the $3D$ O(3) transition.

%Although we chose a specific vison hopping model Eq.~\ref{jh} to
%obtain the vison band structure, the three minima $\vec{Q}_a$ in
%the BZ are stable against any symmetry allowed perturbations on
%Eq.~\ref{jh}. This is because no linear spatial derivative terms
%are allowed in Eq.~\ref{cubic} by transformations
%Eq.~\ref{transforms}.   Thus there are only two ways to
%destabilize the minima in the BZ: ({\it 1}). the current minima
%will be replaced by a new set of minima through a first order
%transition, like the transition between $c-$VBS and $s-$VBS at
%$J/h = 1/8$; ({\it 2}). the sign of the spatial derivative terms
%in Eq.~\ref{jh} changes through a second order Lifshitz
%transition. The second situation will lead to the incommensurate
%VBS order, which will not be discussed in detail here.

\subsubsection{$Z_2$ topological phase and VBS on the square lattices}

The vison dynamics on the square lattice is technically more
complicated than the honeycomb lattice, because in order to solve
the odd $Z_2$ gauge constraint, now the dual quantum Ising model
has to apparently break the lattice symmetry in any specific gauge
choice. The correct lattice symmetry transformation for the dual
vison field $\tau^z$ must be combined with a nontrivial $Z_2$
gauge transformation, $i.e.$ $\tau^z$ carries a projective
representation of the symmetry group. The dual quantum Ising model
has to be invariant under the projective symmetry group (PSG).

\begin{figure}
\begin{center} \includegraphics[width=3.3
in]{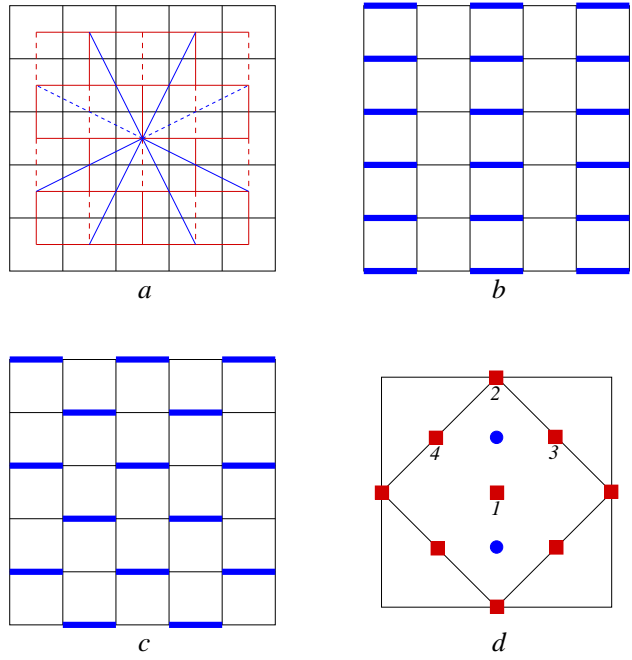}\end{center} \caption{({\it a}), the dual
square lattice. The vison (vortex) hopping on the dashed bonds are
negative. ({\it b}), ({\it c}), the $c-$VBS and $s-$VBS patterns.
({\it d}), the vison (vortex) Brillouin zone. When the nearest
neighbor vison (vortex) hopping is dominant, there are two
inequivalent minima located at $(0, \pm \frac{\pi}{2})$ (circles);
when the 4th neighbor hopping is dominant, there are four
inequivalent minima described by Eq.~\ref{squaresvbsq}.}
\label{squarevison1}
\end{figure}

One of the dual quantum Ising model that is consistent with all
the PSG is \beqn H = \sum_{\bar{p} }- K \tau^x_{\bar{p}} +
\sum_{<\bar{p},\bar{q}>}
J_{\bar{p},\bar{q}}\tau^z_{\bar{p}}\tau^z_{\bar{q}} +
\sum_{\bar{p},\bar{q}}
J^\prime_{\bar{p},\bar{q}}\tau^z_{\bar{p}}\tau^z_{\bar{q}}.
\label{squareising}\eeqn $J$ and $J^\prime$ denote the nearest and
fourth nearest neighbor Ising couplings. $J$ and $J^\prime$ are
chosen to be positive on all the solid bonds, but negative on all
the dashed bonds in Fig.~\ref{squarevison1}$a$.

This quantum Ising model can be analyzed in the same way as the
previous subsection. And with different choices of $J^\prime$ and
$J$ we will find that the $\tau^z$ band structure has multiple
minima in the BZ. The condensate on these minima corresponds to
different VBS pattern. If $J^\prime / J < 0.0858$, there are two
inequivalent minima in the vison band structure, located at
$\vec{Q} = (0, \pm \frac{\pi}{2})$. Again we can expand $\tau^z$
at these two minima as \beqn \tau^z \sim \varphi
e^{i\frac{\pi}{2}y} + \varphi^\ast e^{- i\frac{\pi}{2}y}. \eeqn
The PSG for $\varphi$ reads \beqn \mathrm{T}_x &:& x \rightarrow
x+1, \ \ \varphi \rightarrow e^{i\frac{\pi}{4} x}\varphi^\ast,
\cr\cr \mathrm{T}_y &:& y \rightarrow y + 1, \ \ \varphi
\rightarrow e^{- i\frac{\pi}{4} x}\varphi^\ast, \cr\cr
\mathrm{P}_y &:& x \rightarrow -x, \ \ \varphi \rightarrow
\varphi, \cr\cr \mathrm{P}_x &:& y \rightarrow -y, \ \ \varphi
\rightarrow \varphi, \cr\cr \mathrm{P}_{x+y} &:& x \rightarrow y,
\ y \rightarrow x, \ \ \varphi \rightarrow i\varphi^ \ast. \eeqn
Notice that the reflection $\mathrm{P}_x$ and $\mathrm{P}_y$ are
site-centered reflection of the dual lattice (bond-centered
reflection of the original lattice). The PSG allowed field theory
for $\varphi$ reads \beqn \mathcal{L} = |\partial_\mu \varphi|^2 +
r |\varphi|^2 + g|\varphi|^4 + w (\varphi^8 + \varphi^{\ast 8}).
\label{z8} \eeqn The gauge invariant physical order parameters are
the columnar VBS orders (Fig.~\ref{squarevison1}$b$): \beqn
c-\mathrm{VBS}_x &:& e^{i\frac{\pi}{4}}\varphi^2 + e^{-
i\frac{\pi}{4}}\varphi^{\ast 2}, \cr\cr c-\mathrm{VBS}_y &:& e^{-
i\frac{\pi}{4}}\varphi^2 + e^{i\frac{\pi}{4}}\varphi^{\ast 2}.
\eeqn The quantum phase transition between the $Z_2$ liquid and
the $c-$VBS is a $3D$ XY$^\ast$ transition, since the $Z_8$
anisotropy in Eq.~\ref{z8} is highly irrelevant at the $3D$
XY$^\ast$ fixed point. This result is consistent with previous
studies on fully frustrated Ising model on the cubic lattice
\cite{ffisingcubic1,ffisingcubic2}.

When $J^\prime / J > 0.0858$, the minima of the vison band
structure are shifted to four other inequivalent momenta in the BZ
(Fig.~\ref{squarevison1}$d$): \beqn Q_1 = (0,0), \ \ Q_2 = (0,
\pi), \ \ Q_3 = (\frac{\pi}{2}, \frac{\pi}{2}), \ \ Q_4 = (-
\frac{\pi}{2}, \frac{\pi}{2}). \label{squaresvbsq} \eeqn Notice
all these four modes are real fields, because $Q_a$ are equivalent
to $- Q_a$. Thus these four minima correspond to four different
real fields, which correspond to four different staggered-VBS
state described in Fig.~\ref{squarevison1}$c$.

These analysis can be parallelly generalized to the triangular
lattice. For odd $Z_2$ gauge theory on the triangular lattice, the
dual theory is a frustrated quantum Ising model on the honeycomb
lattice. For the simplest nearest neighbor frustrated quantum
Ising model on the honeycomb lattice, there are four minima in the
vison Brillouin zone, and the low energy field theory of the QCP
between the $Z_2$ liquid and the VBS has a large emergent O(4)
symmetry~\cite{sondhi2001a}, and the liquid-VBS transition belongs
to the 3D O(4)$^\ast$ universality class. The $Z_2$ topological
phase to VBS transition on the Kagome lattice was recently studied
in Ref.~\cite{yehjin}.

\section{Unconventional QCP between ordered phases}

In this section we discuss unconventional QCP between two
different types of ordered phases, $i.e.$ two ordered phases with
different symmetry breaking. More precisely, the GSM of one of the
phases around this QCP should not be the submanifold of the other
phase. When one phase diagram involves two or even more ordered
phases like this, these orders are usually called ``competing
orders". For instance, in the phase diagram of High $T_c$
cuprates, there are both N\'{e}el order and superconductor, as
well as other possible orders such as spin or charge density wave
especially at certain commensurate doping. The classical way of
describing competing orders, is to start with a GL theory that
involves all the relevant competing orders. However, in this
approach it is impossible to get a generic unfine-tuned continuous
quantum phase transition between two different competing orders.
The GL theory will conclude that two competing orders are always
separated by one first order transition, or two (or even more)
continuous transitions.

\begin{figure}
\begin{center} \includegraphics[width=4.1 in]{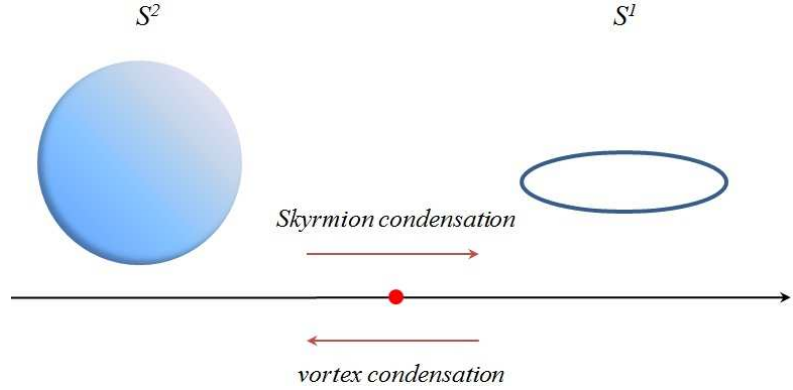}
\end{center}
\caption{The schematic phase diagram of unconventional QCP between
two different competing orders. The phase $A$ has GSM $S^2$, and
it spontaneously breaks symmetry SU(2)$_A$; the phase $B$ with GSM
$S^1$ spontaneously breaks symmetry U(1)$_B$. This QCP can be
interpreted as the condensation of Skyrmions of phase $A$, it can
also be interpreted as condensation of vortices of phase $B$.}
\label{QCP}
\end{figure}

An unfine-tuned direct second order quantum transition between two
competing orders must be an unconventional QCP. The existence of
this unconventional QCP implies that, suppressing one of the
orders necessarily leads to the other order. This effect is
guaranteed when the topological defect of one of the orders
carries the quantum number of the other order. So far, almost all
the unconventional QCP of this type can be {\it roughly} described
with the general formalism described in the following section.

\subsection{General Formalism}

The system has a global symmetry SU(2)$_A$$\times$U(1)$_B$. Phase
$A$ of the phase diagram spontaneously breaks the SU(2)$_A$
symmetry down to U(1)$_A$ symmetry, thus phase $A$ has GSM
$\mathrm{SU(2)}_A/\mathrm{U(1)}_A = S^2$; phase $B$ of the phase
diagram spontaneously breaks the U(1)$_B$ symmetry, thus the phase
$B$ has GSM $S^1$. The Skyrmion defect of phase $A$ carries the
quantum number of U(1)$_B$, thus when this Skrymion condenses, it
not only destroys order $A$, it also induces order $B$. Meanwhile,
the vortex defect of order $B$ carries a fundamental
representation of SU(2)$_A$, thus the condensate of this vortex
not only destroys order $B$, it also leads to phase $A$ that
spontaneously breaks SU(2)$_A$ symmetry.

Now let us describe this QCP from phase $A$. Phase $A$ has GSM
$S^2$, thus it can be described by a unit O(3) vector $\vec{n}$
and the NLSM Eq.~\ref{o3nlsm}. The special property of phase $A$
is that its Skyrmion carries a global U(1)$_B$ symmetry, thus this
Skyrmion is conserved. In order to describe this Skyrmion as a
local excitation instead of a topological defect, it is most
convenient to use the CP(1) field representation: $\vec{n} =
z^\dagger_\alpha \vec{\sigma}_{\alpha\beta} z_{\beta}$. $z_\alpha
= (z_1, z_2)$ is a two-component complex boson, with constraint
$|z_1|^2 + |z_2|^2 = 1$, thus it has three independent degrees of
freedom, while vector $\vec{n}$ only has two degrees of freedom,
thus one of the degrees of freedom of $z_\alpha$ is an unphysical
gauge mode. In fact, $z_\alpha$ is usually parametrized as \beqn
z_\alpha = (z_1, z_2)^t = e^{i\gamma/2}(e^{i\phi/2}
\cos(\theta/2), \ e^{-i\phi/2} \sin(\theta/2))^t, \eeqn here the
angle $\gamma$ is a gauge degree of freedom. Because $z_\alpha$ is
not gauge invariant, if we describe phase $A$ using $z_\alpha$,
then $z_\alpha$ is automatically coupled to a U(1) gauge field
$a_\mu$, and the field theory that describes phase $A$ reads \beqn
\mathcal{L} = \sum_\alpha |(\partial_\mu - i a_\mu)z_\alpha|^2 +
r|z_\alpha|^2 + u (\sum_\alpha |z_\alpha|^2)^2 +
\frac{1}{e^2}(f_{\mu\nu})^2. \label{cp1} \eeqn In this equation we
have softened the constraint $|z_1|^2 + |z_2|^2 = 1$, whose effect
has been replaced by the interaction $u$ term.

This CP(1) representation has a great advantage: the Skyrmion of
the vector $\vec{n}$ is precisely the flux quantum of $a_\mu$:
\beqn \frac{1}{8\pi} \epsilon_{abc} \epsilon_{ij} n^a
\partial_i n^b \partial_j n^c = \frac{1}{2\pi} \epsilon_{ij} \partial_i a_j.
\eeqn Because the Skyrmion carries a global U(1)$_B$ symmetry, the
Skyrmion is conserved, thus when it condenses the system will have
a Goldstone mode. Because the Skyrmion is mapped to the U(1) gauge
flux quantum, this Skyrmion Goldstone mode is precisely dual to
the photon excitation of $a_\mu$.

In terms of the CP(1) field theory, this phase diagram is
interpreted as follows: phase $A$ is the condensate of CP(1) field
$z_\alpha$, and the gauge field $a_\mu$ is gapped due to the Higgs
mechanism. Phase $A$ has GSM $S^2$ characterized by vector
$z^\dagger \vec{\sigma} z$. Phase $B$ is the gapped phase of
$z_\alpha$, and in this phase the gauge field $a_\mu$ is in its
photon phase, which is precisely the condensate of its gauge flux
(the duality discussed in section {\bf 2}). Because the gauge flux
carries the U(1)$_B$ quantum number, this photon phase
spontaneously breaks the U(1)$_B$ symmetry, and it has GSM $S^1$.

We can also understand this QCP from phase $B$. Phase $B$ has GSM
$S^1$, which is equivalent to a superfluid phase. In section {\bf
2}, we derived the dual description of the SF phase, which is the
bosonic QED Eq.~\ref{sfdual}, where the vortex of the SF phase is
described by a bosonic scalar field $\Phi$ that couples to the
dual U(1) gauge field $a_\mu$. In the current case, since we
assumed that the vortex of phase $B$ carries a fundamental
representation of the SU(2)$_A$ symmetry, then the dual theory
actually becomes precisely the CP(1) field theory Eq.~\ref{cp1},
and the CP(1) field $z_\alpha$ precisely corresponds to the vortex
of phase $B$.

The phase transition between phase $A$ and $B$ is described by
``fractionalized particles" $z_\alpha$ instead of physical order
parameters, thus this type of QCP is called ``deconfined QCP".
Although the CP(1) field theory was ``derived" from the O(3)
model, the universality class of the QCP described by
Eq.~\ref{cp1} is {\it very different} from the O(3) Wilson-Fisher
fixed point. The O(3) Wilson-Fisher quantum phase transition is
sandwiched between a phase with GSM $S^2$ and a fully gapped
trivial disordered phase, which is very different from phase $B$.
In fact, the O(3) Wilson-Fisher fixed point is equivalent to the
case where the Skyrmion (the gauge flux) is {\it not} conserved,
$i.e.$ the U(1)$_B$ symmetry is absent. In the language of the
CP(1) model, an unconserved Skyrmion number corresponds to an {\it
unconserved} U(1) gauge flux of $a_\mu$, thus the flux condensate
has no Goldstone mode, $i.e.$ the photon excitation is fully
gapped. The U(1) gauge field with unconserved flux is precisely
the compact U(1) gauge theory. Thus the O(3) Wilson-Fisher
universality class is equivalent to a $compact$-CP(1) model, while
the deconfined QCP is described by a $noncompact$-CP(1) model.

\subsection{Examples of deconfined QCPs}

In this subsection we discuss two examples of deconfined QCPs.

The first example of deconfined QCP that was discussed is the
N\'{e}el-VBS transition of spin-1/2 quantum magnet on the square
lattice: the phase $A$ is the AF N\'{e}el order that breaks the
spin rotation SU(2) symmetry, while phase $B$ is the VBS phase
that only breaks lattice translation and rotation
symmetry~\cite{deconfine1,deconfine2}. It appears that this
transition is different from the general case discussed in the
previous subsection, since the VBS phase breaks a discrete four
fold rotation lattice symmetry, instead of a continuous U(1)$_B$
symmetry. However, there is a strong analytical and numerical
evidence which suggests that the discrete four fold rotation
symmetry is enlarged to a continuous U(1)$_B$ symmetry at the
QCP~\cite{sandvik1,ribhu,sandvik2}. Thus the GSM of phase $B$ is
enlarged to $S^1$ close to the QCP.

The essence of the deconfined QCP is the physical quantum number
carried by topological defects in both phase $A$ and $B$. It was
shown by Haldane and Sachdev that the Skyrmion of N\'{e}el order
carries lattice momentum, thus when the Skyrmion of N\'{e}el order
condenses, it spontaneously breaks the lattice symmetry, $i.e.$
the system automatically enters the VBS
order~\cite{haldane1988,sachdev1990}. Later on Senthil and Levin
also demonstrated that the discrete $Z_4$ vortex of the VBS order
carries a spin-1/2 spinon, thus as long as the $Z_4$ rotation
symmetry of the lattice is enlarged to U(1)$_B$ symmetry at the
QCP, this QCP is exactly equivalent to the general formalism
discussed in the previous subsection~\cite{levinsenthil}.

In Ref.~\cite{grover}, the authors proposed another deconfined
QCP. In this phase diagram, phase $A$ is a quantum spin Hall
insulator on the honeycomb lattice, but the quantum spin Hall
(QSH) state is generated by spontaneously breaking the spin
symmetry, while preserving the time-reversal symmetry, thus the
QSH state discussed in Ref.~\cite{grover} has GSM $S^2$, which is
equivalent to phase $A$ in the general formalism. Also, it was
demonstrated in Ref.~\cite{abanov2000} that the Skyrmion of the
QSH vector carries charge-$2e$. Thus the Skyrmion of the QSH
vector is conserved, and if this Skyrmion condenses, the system
enters a $s-$wave superconductor. The transition between the QSH
and $s-$wave SC is precisely described by Eq.~\ref{cp1}. In this
model, SU(2)$_A$ is the spin SU(2) symmetry, while the U(1)$_B$ is
the charge U(1) symmetry.

\section{Global phase diagram of unconventional QCPs}

So far in our paper we have discussed three types of
unconventional QCPs:

$ $

1. QCP between $Z_2$ topological order and spin ordered phase;

$ $

2. QCP between $Z_2$ topological order and VBS phase;

$ $

3. Deconfined QCP between two different ordered phases that
spontaneously break two different symmetries.

$ $

In this section we will discuss a single unified theory that
contains all these phenomena in one phase diagram. This theory was
introduced in Ref.~\cite{xusachdev} and Ref.~\cite{xu2010}, and it
was applied to different microscopic systems. Before we discuss
the physical motivation of this theory, let us first write down
the Lagrangian of the unified field theory: \beqn \mathcal{L} &=&
\sum_{\alpha = 1}^{N_z} |(\partial_\mu - i a_\mu) z_\alpha|^2 +
s_z |z_\alpha|^2 + \sum_{\alpha = 1}^{N_v} |(\partial_\mu - i
b_\mu) v_\alpha|^2 + s_v |v_\alpha|^2  \cr\cr &+&
\frac{i}{\pi}\epsilon_{\mu\nu\rho} a_\mu
\partial_\nu b_\rho + \cdots \label{mcs}\eeqn In this field theory, there
are two types of matter fields, $z_\alpha$ and $v_\alpha$, and
they are interacting with each other through a mutual Chern-Simons
theory, which grants them a mutual semion statistics $i.e.$ when
$v_\alpha$ adiabatically encircles $z_\alpha$ through a closed
loop, the system wave-function acquires a minus sign.

The field theory Eq.~\ref{mcs} has symmetry
SU($N_z$)$\times$SU($N_v$). However, depending on the details of
the microscopic model, the higher order interactions between
matter fields can break this symmetry down to its subgroups. We
will first ignore this high order symmetry breaking effects, and
focus on the case with $N_z = 2$, and $N_v = 1$.

In Ref.~\cite{xusachdev}, the authors used the model Eq.~\ref{mcs}
with $N_z =2$, $N_v = 1$ to describe the global phase diagram of
spin-1/2 quantum magnets on a distorted triangular lattice, which
is a very common structure in many materials. Here $z_\alpha$ is a
bosonic spin-1/2 spinon, and $v$ is the low energy mode of vison,
which is a complex scalar field, like the complex $\psi$ and
$\varphi$ field introduced in Eq.~\ref{complex1} and
Eq.~\ref{complex2}. The phase diagram of this model is tuned by
two parameters: $s_z$ and $s_v$, and depending on the sign of
these two parameters, there are in total four different phases
(Fig.~\ref{QCP}):

\begin{figure}
\begin{center} \includegraphics[width=3.5 in]{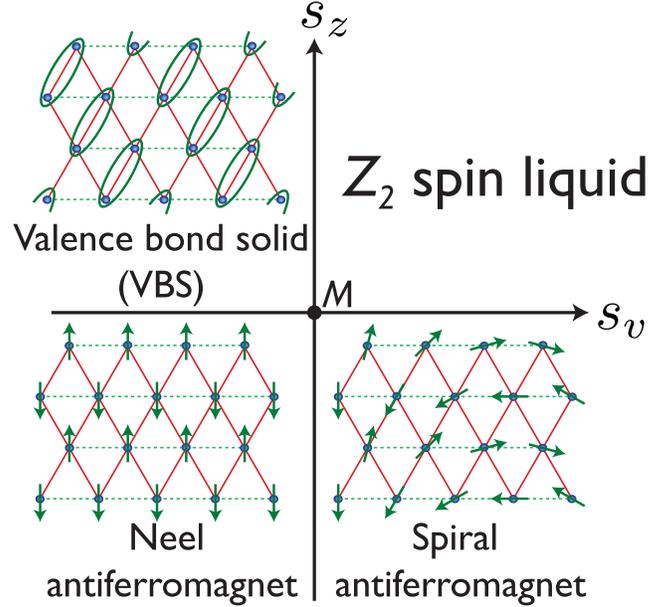}
\end{center}
\caption{The global phase diagram of Eq.~\ref{mcs}, which
describes four different standard spin states on a distorted
triangular lattice.} \label{QCP}
\end{figure}

{\it Phase 1.} This is the phase with $s_z > 0$, $s_v > 0$. In
this phase, both matter fields $z_\alpha$ and $v$ are gapped, and
they have a topological statistic interaction through the mutual
CS theory. Since all the matter fields are gapped, the low energy
properties of phase 1 is described by the mutual CS theory only.
Thus phase 1 is the $Z_2$ topological phase, described by $Z_2$
gauge theory Eq.~\ref{z2gauge1}.

%If we solve the mutual Chern-Simons theory on the torus, we will
%see that its ground state is four-fold degenerate, thus this phase
%is precisely a gapped $Z_2$ topological phase $i.e.$ a deconfined
%phase of a $Z_2$ gauge field.

{\it Phase 2.} $s_v > 0$, $s_z < 0$. When $N_z = 2$, this phase
corresponds to a condensate of CP(1) field $z_\alpha = (z_1, z_2)$
while coupling to a $Z_2$ gauge field, thus this phase has GSM
SO(3). Physically this phase corresponds to the incommensurate
spiral SDW.

{\it Phase 3.} $s_v < 0$, $s_z > 0$. This is a phase where $v$
condenses while $z_\alpha$ is gapped out. This phase is the VBS
phase that breaks the reflection and translation symmetry of the
lattice.

{\it Phase 4.} $s_v < 0$, $s_z < 0$. This is a phase where both
$z_\alpha$ and $v$ condense, and a careful analysis will conclude
that this is precisely the collinear N\'{e}el phase with GSM
$S^2$.

According to the unified theory Eq.~\ref{mcs}, the QCP between
phase 1 and 2 ($Z_2$ topological phase and spiral SDW) is the 3D
O(4)$^\ast$ transition that was described by Eq.~\ref{o4field};
The QCP between phase 1 and 3 ($Z_2$ topological phase and VBS) is
the 3D XY$^\ast$ transition described by Eq.~\ref{z8}. The QCP
between phase 3 and 4 is the deconfined QCP that is described by
the noncompact CP(1) field theory Eq.~\ref{cp1}. A more detailed
discussion of the phases and QCPs of Eq.~\ref{mcs} can be found in
Ref.~\cite{xusachdev}.

All of these phases have been observed in real frustrated quantum
magnets on the (distorted) triangular lattice. For example, a
noncollinear spiral SDW was observed in
$\mathrm{Cs_2CuCl_4}$~\cite{cscucl}; spin liquid phases were
discovered in $\kappa-\mathrm{(ET)_2 Cu_2
(CN)_3}$~\cite{kanoda0,kanoda1,kanoda2,myamashita,kanoda3},
$\mathrm{EtMe_3Sb[Pd(dmit)_2]_2}$~\cite{kato2,kato3,kato4,kato5,kato6,kato7},
$\mathrm{Ba_3CuSb_2O_9}$~\cite{cuspinliquid}, and
$\mathrm{Ba_3NiSb_2O_9}$~\cite{nispinliquid}; a VBS phase was
observed in
$\mathrm{(C_2H_5)(CH_3)_3P[Pd(dmit)_2]_2}$~\cite{kato4}; and many
materials that belong to the dmit family have collinear N\'{e}el
order at low temperature. All of these phases, including the QCPs
between them can be unified using one single Lagrangian
Eq.~\ref{mcs}.

In Ref.~\cite{xu2010}, Eq.~\ref{mcs} was used to describe the
phase diagram of the Hubbard model on the honeycomb lattice.
$z_\alpha$ and $v_\alpha$ are the fundamental excitations of spin
SU(2) and charge SU(2) symmetry of the Hubbard model at
half-filling. Since the maximal symmetry of interacting electron
systems is SO(4)$\sim (\mathrm{SU(2)_{spin} \times
SU(2)_{charge}})/Z_2$, there is an extra factor of $Z_2$ in the
GSM of all the phases in this phase diagram. For example, here the
phase with $s_z > 0$ and $s_v > 0$ is a $Z_2 \times Z_2$
topological phase, and Ref.~\cite{xu2010} identified this phase as
the fully gapped spin liquid phase observed by quantum Monde Carlo
simulation on the Hubbard model on a honeycomb
lattice~\cite{meng}.

\section{Summary and Extentions}

So far we have discussed unconventional QCPs around topological
phases, and the QCPs between competing orders. However, this
discussion is far from being general. We have a more or less
complete understanding about QCPs around the $Z_2$ topological
phase, and it is straightforward to generalize this understanding
to $Z_N$ topological phases. However, the QCPs around other
topological phases are less understood. One major limitation of
our description is that, the physical picture of the QCPs
discussed so far all relies on ``condensation" of certain bosonic
point particles. But there is no reason to believe this picture
can be applied to all the QCPs in strongly interacting many-body
systems. For example, a large class of topological phases can be
described using loop or string like variables, instead of point
particles~\cite{wangfreedman,levinwen}. Some of the phases
described by loop variables have a dual description in terms of
point particles~\cite{levingu}, thus the formalism described in
this paper may still apply, but we do not have a general formalism
to describe QCPs driven by extended objects.

Another complication of topological phases is that, their low
energy excitations can carry nontrivial anyonic or even nonabelian
statistics. Some of these excitations can be described as bosons
coupled to a Chern-Simons field, but a more general and complete
formalism of dealing with particles with nontrivial statistics is
still demanded. The condensation of anyons with nontrivial
statistics usually drives the system into a different topological
phases, and in this case the two states around the QCP has the
same symmetry. Examples of quantum critical points between
different topological orders have been studied in
Ref.~\cite{wenqcp1,wenqcp2,wenqcp3,wenqcp4}.

Unconventional QCPs are very easy to detect experimentally,
because of its large anomalous dimension associated with physical
order parameters. For example, in 2+1 dimension, if there is a QCP
between a magnetic ordered phase and a disordered phase, then in
the quantum critical regime with finite temperature the NMR
relaxation rate $1/T_1$ has the universal scaling $1/T_1 \sim
T^{\eta}$. Thus the unusual anomalous dimension $\eta$ can be
probed conveniently in experiments. In Ref.~\cite{kanoda1}, it was
reported that the NMR relaxation rate of material
$\kappa-\mathrm{(ET)_2Cu_2(CN)_3}$ scales as $1/T_1 \sim T^{a}$,
where $a \sim 1.5$. This is qualitatively consistent with the 3D
XY$^\ast$ and O(4)$^\ast$ QCP discussed in section {\bf 3}. This
observation led to the conjecture that the organic material
$\kappa-\mathrm{(ET)_2Cu_2(CN)_3}$ is close to an unconventional
QCP~\cite{xuqi}.

Unconventional quantum critical point is a rapidly developing
field, and it is impossible for us to review every related topic.
Besides the subjects included in this paper, there are a few other
types of exotic QCPs that are beyond the Landau's paradigm. For
example, our paper has focused on the unconventional QCPs in two
spatial dimensions, while the idea of deconfined QCPs has been
generalized to three dimensional lattices as
well~\cite{motrunich,nussinov}. Another special type of exotic QCP
in fermionic systems was reviewed in Ref.~\cite{russian}.
Unconventional phase transitions at finite temperature in
classical systems have also been discussed in special models, for
example the classical dimer models on three dimensional lattices
\cite{powell1,powell2,balentsdimer}.

%\begin{acknowledgement}

%\begin{acknowledgements}

$ $

{\bf Acknowledgement}

The author is supported by the Sloan Foundation.

%\end{acknowledgements}

%\end{acknowledgement}

%However, for a general topological phase, the low energy
%excitations can have much more exotic statistics, then we now
%longer have a well-understood formalism.

%\bibliographystyle{ws-ijmpb}

%\bibliography{References}

\end{document}